\documentclass[final,3p,times,twocolumn]{elsarticle}

\usepackage{amssymb}

\usepackage{wallpaper}
\usepackage{dcolumn}
\usepackage{bm}
\usepackage{pdfpages}
\usepackage{graphics}
\usepackage[colorlinks,plainpages=false]{hyperref}
\usepackage{bbding}

\begin{document}
\begin{frontmatter}



\title{Quark condensate and chiral symmetry restoration in neutron stars}

\author[SunAdr1,XiaAdr1]{Hao-Miao~Jin}

\author[XiaAdr1]{Cheng-Jun Xia\corref{cor1}}
\ead{cjxia@nit.zju.edu.cn}

\author[SunAdr1]{Ting-Ting~Sun\corref{cor1}}
\ead{ttsunphy@zzu.edu.cn}

\author[PengAdr1,PengAdr2,PengAdr3]{Guang-Xiong~Peng\corref{cor1}}
\ead{gxpeng@ucas.ac.cn}


\cortext[cor1]{Corresponding author.}

\address[SunAdr1]{School of Physics and Microelectronics, Zhengzhou University, Zhengzhou 450001, China}
\address[XiaAdr1]{School of Information Science and Engineering, NingboTech University, Ningbo 315100, China}
\address[PengAdr1]{School of Nuclear Science and Technology, University of Chinese Academy of Sciences, Beijing 100049, China}
\address[PengAdr2]{Theoretical Physics Center for Science Facilities, Institute of High Energy Physics, P.O. Box 918, Beijing 100049, China}
\address[PengAdr3]{Synergetic Innovation Center for Quantum Effects and Application, Hunan Normal University, Changsha 410081, China}

\begin{abstract}
Based on an equivparticle model, we investigate the in-medium quark condensate in neutron stars. Carrying out a Taylor expansion of the nuclear binding energy to the order of $\rho^3$, we obtain a series of EOSs for neutron star matter, which are confronted with the latest nuclear and astrophysical constraints. The in-medium quark condensate is then extracted from the constrained properties of neutron star matter, which decreases non-linearly with density. However, the chiral symmetry is only partially restored with non-vanishing quark condensates, which may vanish at a density that is out of reach for neutron stars.
\end{abstract}


\end{frontmatter}

\section{\label{sec:intro}Introduction}

Significant progresses were made in understanding strongly interacting matter at large temperatures, where a smooth crossover from hadronic matter (HM) to quark gluon plasma (QGP) were observed~\cite{Borsanyi2014_PLB730-99, Bazavov2014_PRD90-094503}. The state of matter at large densities, however, is still veiled in mystery due to the difficulties in lattice QCD simulations. It is thus essential for us to investigate the properties of dense matter with both nuclear and astrophysical constraints.

According to various investigations on the structures and reactions of finite nuclei, the properties of nuclear matter around the saturation density $\rho_{0} = 0.16$ fm$^{-3}$ are well constrained with the binding energy $\epsilon_{0}(\rho _{0})\approx -16$ MeV, the incompressibility $K = 240 \pm 20$ MeV~\cite{Shlomo2006_EPJA30-23}, the symmetry energy $S(\rho_{0}) = 31.7 \pm 3.2$ MeV and its slope $L = 58.7 \pm 28.1$ MeV~\cite{Li2013_PLB727-276, Oertel2017_RMP89-015007}. Particularly, at about two thirds of the nuclear saturation density $\rho_\mathrm{on}=0.1\ \mathrm{fm}^{-3}$, the symmetry energy is fixed accurately according to finite nuclei properties, i.e., $S(\rho_\mathrm{on}) =25.5 \pm 1.0$ MeV~\cite{Centelles2009_PRL102-122502, Brown2013_PRL111-232502}. This indicates a relation between the symmetry energy and its slope~\cite{Horowitz2001_PRL86-5647}
\begin{equation}
  S(\rho_{0}) \approx 26\ \mathrm{MeV} + \frac{L}{9}. \label{Eq:SL}
\end{equation}
Meanwhile, the slope of symmetry energy was shown to be linearly correlated with the neutron skin thickness $\Delta R_{np}$~\cite{Zhang2013_PLB726-234}. Based on the recent measurements of $\Delta R_{np}$ in ${}^{208}$Pb, the pioneering Lead Radius Experiment (PREX) II suggests $L = 106 \pm 37$ MeV~\cite{PREX2021_PRL126-172502, Reed2021_PRL126-172503}, while higher accuracy is expected in the upcoming Mainz Radius Experiment. For nuclear matter at larger densities, the experimental studies with heavy ion collisions also provide important constraints, e.g., those in Refs.~\cite{Danielewicz2002_Science298-1592, Liu2021_PRC103-014616}.

Being one of the most dense objects in the Universe, pulsar-like compact stars provide natural laboratories for dense matter, where the density may reach $\sim$$8\rho_0$. The precise mass measurements of PSR J0348+0432 ($2.01 \pm 0.04\ M_\odot$)~\cite{Antoniadis2013_Science340-1233232} and PSR J0740+6620 ($2.14^{+0.10}_{-0.09}M_{\odot}$)~\cite{Cromartie2020_NA4-72} have put strong constraints on the equation of state (EOS) of supranuclear dense matter. By analyzing the gravitational waves emitted from the binary neutron star merger event GW170817, the tidal deformability of $1.4 M_{\odot}$ neutron star are constrained within $70\leq \Lambda_{1.4}\leq 580$, corresponding to the radii $11.9^{+1.4}_{-1.4}$ km~\cite{LVC2018_PRL121-161101}. Based on pulse-profile modeling~\cite{Watts2019_SCPMA62-29503}, both the mass and radius of a neutron star can be measured, e.g., the observations of PSR J0030+0451 and PSR J0740+6620 have placed the radii of $1.4 M_{\odot}$ and $2.08 M_{\odot}$ neutron stars at $12.45\pm 0.65$ km and $12.35\pm 0.75$ km (68\% credible region)~\cite{Miller_2021qha}. Combining all those observations, the most stringent constraint on the EOSs of dense matter can be obtained~\cite{Li2020, Zhang2020_PRC101-034303}.

At large enough densities, it is expected that HM undergoes a deconfinement phase transition and forms quark matter (QM). A recent investigation adopting various constraints has found that neutron star matter at highest densities exhibits characteristics of QM~\cite{Annala2020_NP}, while the simultaneous mass-radius measurements of PSR J0030+0451 and PSR J0740+6620 have ruled out any strong first-order phase transitions at the center of neutron stars with masses between $\sim$$1.4 M_{\odot}$ and $\sim$$2.08 M_{\odot}$~\cite{Miller_2021qha}. In such cases, if deconfinement phase transition is of first-order, there should be quark-hadron mixed phase inside hybrid stars~\cite{Glendenning2000, Heiselberg1993_PRL70-1355, Voskresensky2002_PLB541-93, Tatsumi2003_NPA718-359, Endo2005_NPA749-333, Maruyama2007_PRD76-123015, Peng2008_PRC77-065807, Klahn2013_PRD88-085001, Yasutake2014_PRC89-065803, Bombaci2016_IJMPD-1730004, Xia2019_PRD99-103017, Maslov2019_PRC100-025802, Xia2020_PRD102-023031}. Meanwhile, it was shown that a smooth crossover from HM to QM can also accommodate the stringent constraints from pulsar observations~\cite{Masuda2013_ApJ764-12, Zhao2015_PRD92-054012, Kojo2015_PRD91-045003, Whittenbury2016_PRC93-035807, Bai2018_PRD97-023018, Baym2019_ApJ885-42}.

Particularly, the crossover between HM and QM is expected to be bridged by quarkyonic matter~\cite{Fukushima2016_ApJ817-180}. The quarkyonic phase was initially proposed by McLerran and Pisarski in the large $N_c$ limit~\cite{McLerran2007_NPA796-83}, which is comprised of ``a quark Fermi sea" and ``a baryonic Fermi surface". For quarks near the Fermi surface, the active degrees of freedom are still baryons. Under such considerations, recently McLerran and Reddy constructed explicitly the quarkyonic matter which naturally explains the observed properties of neutron stars~\cite{McLerran2019_PRL122-122701}. Based on an extended Nambu-Jona-Lasinio (NJL) model with Polyakov loops, McLerran, Redlich, and Sasaki found  that the quarkyonic transition is indeed a crossover at $N_c=3$~\cite{McLerran2009_NPA824-86}. The baryon density changes very rapidly at the quarkyonic transition, and one may expect that the quarkyonic and chiral phase transitions are entangled. It is thus necessary to investigate the in-medium quark condensate inside neutron stars, which was recently carried out adopting a parity doublet model for nuclear matter, NJL model for quark matter, and an interpolation at the intermediate density~\cite{Minamikawa2021}.

In this work we study the in-medium quark condensate systematically based on an equivparticle model~\cite{Peng2002_PLB548-189}. Since the hadron resonance gas (HRG) model well describes the crossover from HM to QGP at large temperatures~\cite{Karsch2003_PLB571-67, Andronic2018_Nature561-321}, we expect that the crossover from HM to QM can be described with baryonic degrees of freedom alone. In particular, we obtain a series of EOSs of nuclear matter by carrying out a Taylor expansion to the order of $\rho^3$~\cite{Zhang2018_ApJ859-90}, which are then confronted with both nuclear and astrophysical constraints. For those consistent with various constraints, we extract the in-medium quark condensate in the framework of equivparticle model~\cite{Peng2002_PLB548-189}, which bypasses the extra difficulty in taking a derivative of the system energy density with respect to the quark current mass. It is found that in most cases, the in-medium quark condensate in neutron stars decreases with density. However, the chiral symmetry is only partially restored with non-vanishing quark condensates, which are consistent with the findings in Ref.~\cite{Minamikawa2021}.

This paper is organized as follows. The theoretical framework in obtaining the EOSs of neutron star matter and the corresponding in-medium quark condensate are presented in Section~\ref{sec:the}. Section~\ref{sec:num} is devoted to the discussion of the numerical results, where the in-medium quark condensate in neutron stars that are consist with observations are obtained. Finally, a summary is given in Section~\ref{sec:con}.

\section{\label{sec:the}Theoretical framework}
\subsection{\label{sec:the_NS}Nuclear matter and neutron stars}
The binding energy per nucleon for nuclear matter at given density $\rho=\rho_n + \rho_p$ and isospin asymmetry $\delta =(\rho_n - \rho_p)/\rho$ can be obtained approximately with
\begin{equation}
\epsilon(\rho, \delta)\approx \epsilon_{0}(\rho)+S(\rho)\delta ^{2}, \label{Eq:Et}
\end{equation}
where $\epsilon_{0}(\rho)$ is the binding energy in symmetric nuclear matter (SNM) and $S(\rho)$ the symmetry energy. Expanding $\epsilon_{0}(\rho)$ and $S(\rho)$ in Taylor series and omitting higher order terms, we have
\begin{eqnarray}
\epsilon_{0} &=&  \epsilon_0(\rho_0)+\frac{K}{18} x^{2}+\frac{J}{162} x^{3}, \label{Eq:E0}\\
S     &=&   S(\rho_0) +\frac{L}{3} x +\frac{K_\mathrm{sym}}{18}x^{2}  + \frac{J_\mathrm{sym}}{162}x^{3}, \label{Eq:Es}
\end{eqnarray}
with $x\equiv\left({\rho}/{\rho_0} -1\right)$, $\rho_{0} = 0.16$ fm$^{-3}$, $\epsilon_{0}(\rho _{0}) \approx -16$ MeV, and $S(\rho_{0}) = 31.7 \pm 3.2$ MeV. The coefficients of the higher order terms in Eq.~(\ref{Eq:E0}) are the incompressibility $K$ and skewness $J$ of SNM, while those in Eq.~(\ref{Eq:Es}) are the slope $L$, curvature $K_\mathrm{sym}$, and skewness $J_\mathrm{sym}$ of the symmetric energy. According to various pulsar observations, a recent study has found that $L = 57.7\pm 19$ MeV and $K_\mathrm{sym}= -107 \pm 88$ MeV~\cite{Li2021_Universe7-182}. It is worth mentioning that the expansions in Eqs.~(\ref{Eq:E0}) and (\ref{Eq:Es}) are not converging at suprasaturation densities, which can be resolved adopting other expansion techniques~\cite{Margueron2018_PRC97-025805, Cai2021_PRC103-054611}.

For given coefficients in Eqs.~(\ref{Eq:E0}) and (\ref{Eq:Es}), the energy density of nuclear matter can be obtained with
\begin{equation}
E_\mathrm{b}(\rho, \delta)=\rho \epsilon(\rho, \delta) +\rho M_{N}, \label{Eq:E_expand}
\end{equation}
where the binding energy $\epsilon(\rho, \delta)$ is fixed by Eq.~(\ref{Eq:Et}) and $M_{N}=938$ MeV is the rest mass of nucleons.
To obtain the EOSs of neutron star matter, the contributions of leptons should be considered, where the total energy density $E(\rho, \delta)$ of $npe\mu$ matter reads
\begin{equation}
  E(\rho, \delta) =E_\mathrm{b}(\rho, \delta) + \sum_{l=e,\mu}E_{l}(\rho, \delta).
\end{equation}
Here $E_l (\rho, \delta)$ is the energy density of leptons, which is determined by
\begin{equation}
E_{l}(\rho, \delta)=\frac{m_l^{4}}{8\pi ^2} f\left(\frac{\sqrt[3]{3\pi^2 \rho_l}}{m_l}\right),
 \end{equation}
with the electron mass $m_e=0.511$ MeV, muon mass $m_\mu=105.66$ MeV, and
\begin{equation}
  f(y)\equiv \left[y\sqrt{1+y^2}\left(1+2y^2\right)-\mathrm{arcsh}(y)\right]. \label{Eq:fx}
\end{equation}
The chemical potential of particle type $i$ can then be calculated from
\begin{equation}
\mu_i=\frac{\partial E(\rho, \delta)}{\partial \rho_i}.
\end{equation}
Through the $\beta$-equilibrium condition $\mu _{e}=\mu_{\mu }=\mu_{n}-\mu_{p}$ and the charge neutrality condition $\rho _{p}=\rho _{e}+\rho_\mu$, we can obtain the isospin asymmetry $\delta(\rho)$ and relative particle fractions ($\rho_i/\rho$ with $i=p,n,e,\mu$) of neutron star matter at fixed density $\rho$. The pressure of the system can be evaluated with
\begin{equation}
  P(\rho, \delta) = \sum_i \rho_i \mu_{i} -E(\rho ,\delta ).
 \end{equation}

\subsection{\label{sec:the_Equiv}Equivparticle model on quark condensate}
In this work, we use an equivparticle model to extract the quark condensate of neutron star matter~\cite{Peng1999_PRC61-015201, Peng2002_PLB548-189, Wen2005_PRC72-015204, Xia2014_PRD89-105027}. The basic idea of equivparticle model is to define an equivalent Hamiltonian density with a variable quark mass as
\begin{equation}
 H_\mathrm{eqv}=H_{K}+\sum_{i}m_{i}\bar{q_{i}}q_{i}, \label{Eq:Hequiv}
\end{equation}
where $H_{K}$ represents the kinetic term and $m_{i}$ the equivalent mass of quark $i$. At the same time, the QCD Hamiltonian density can be schematically written as
\begin{equation}
 H_\mathrm{QCD}=H_{K}+\sum_{i}m_{0i}\bar{q_{i}}q_{i} + H_\mathrm{I}, \label{Eq:HQCD}
\end{equation}
where $m_{i0}$ is the quark current mass and $H_\mathrm{I}$ corresponds to the interaction part. In order for $H_\mathrm{eqv}$ to reflect the characteristics of the original QCD system, the equivalent mass $m_i$ needs to be fixed by fulfilling
\begin{equation}
\langle H_\mathrm{eqv} \rangle_\rho - \langle H_\mathrm{eqv} \rangle_0 = \langle H_\mathrm{QCD} \rangle_\rho - \langle H_\mathrm{QCD} \rangle_0. \label{Eq:Heq}
\end{equation}
Here $\langle O \rangle_\rho \equiv \langle\rho | O |\rho \rangle$ represents the expectation value of operator $O$ in neutron star matter with density $\rho$, which needs to be subtracted by the vacuum contribution $\langle O \rangle_0 \equiv \langle 0 | O | 0 \rangle$. The equivalent mass $m_{i}$ can be determined by
\begin{equation}
 m_{i}=m_{i0}+m_\mathrm{I}, \label{Eq:mass}
\end{equation}
where $m_{i0}$ is the quark current mass and $m_\mathrm{I}$ accounts for the strong interaction among quarks. Substituting Eqs.~(\ref{Eq:Hequiv}) and (\ref{Eq:HQCD}) into Eq.~(\ref{Eq:Heq}), we then obtain the interacting part of the equivalent mass
\begin{equation}
  m_\mathrm{I}=\frac{E_\mathrm{I}}{\Sigma_{i}(\langle\bar{q_{i}}q_{i}\rangle_\rho - \langle\bar{q_{i}}q_{i}\rangle_{0})}, \label{Eq:massI}
\end{equation}
with the interacting energy density
\begin{equation}
 E_\mathrm{I}=\langle H_\mathrm{I} \rangle_\rho -\langle H_\mathrm{I}\rangle_0.
\end{equation}
Note that in obtaining Eq.~(\ref{Eq:massI}) we have assumed a uniformly distributed quark condensate, i.e., neglecting any local fluctuations inside nucleons by taking the spatial average, which gives $\langle m_{i}\bar{q_{i}}q_{i}\rangle = m_{i}\langle\bar{q_{i}}q_{i}\rangle$.

Since nuclear matter only have two quark flavors, i.e., the up ($u$) and down ($d$) quarks whose current masses are approximately equal to each other, we can then assume an exact isospin symmetry, where $m_{u0}=m_{d0}\equiv m_{0}$, $m_{u}=m_{d}\equiv m$, $\langle\bar{u}u \rangle_0 = \langle\bar{d}d \rangle_0 \equiv \langle\bar{q}q\rangle_\rho$, and $\langle\bar{u}u \rangle_\rho + \langle\bar{d}d \rangle_\rho \equiv 2 \langle\bar{q}q\rangle_\rho$. To calculate the quark condensate, we rewrite Eq.~(\ref{Eq:massI}) and obtain
\begin{equation}
 \frac{\langle\bar{q}q\rangle_\rho} {\langle\bar{q}q\rangle_0}=1-\frac{1}{3n^*}\frac{E_\mathrm{I}}{m_\mathrm{I}}. \label{Eq:cond}
\end{equation}
The vacuum quark condensate $\langle\bar{q}q\rangle_0$ as well as the chiral restoration density $n^*$ in the model independent linear expression~\cite{Cohen1992_PRC45-1881} can be obtained according to the GellMan-Oakes-Renner relation $-2m_{0}\langle\bar{q}q\rangle_{0}=m_{\pi}^{2}f_{\pi}^{2}$~\cite{Gell-Mann1968_PR175-2195}, which gives
\begin{equation}
   n^{*}=-\frac{2}{3}\langle\bar{q}q\rangle_{0}=\frac{m_{\pi}^{2}f_{\pi}^{2}}{3m_0} = 0.985\ \mathrm{fm}^{-3}
\end{equation}
with $m_{\pi}\approx 140$ MeV being the pion mass, $f_{\pi}\approx 93.2$ MeV the pion decay constant, and $m_0=7.5$ MeV the average current mass of light quarks.

Due to the fact that $H_\mathrm{eqv}$ has the same form of a free system with the equivalent particle mass $m_{i}$, the energy density can then be obtained with
\begin{equation}
E(\rho_u, \rho_d, m) = \frac{m^{4}g}{16\pi^{2}} \sum_{q=u,d} f\left(\frac{\sqrt[3]{6\rho_q\pi^2}}{m\sqrt[3]{g}}\right). \label{Eq:E_equiv}
\end{equation}
Here $f(y)$ is given by Eq.~(\ref{Eq:fx}) and $g = 2 ($spin$) \times 3 ($color$) =6$ the degeneracy factor of quarks. To fix the equivalent mass $m$, we need to reproduce the energy density fixed by Eq.~(\ref{Eq:E_expand}) with Eq.~(\ref{Eq:E_equiv}), i.e.,
\begin{equation}
 E(\rho_u, \rho_d, m) = E_\mathrm{b}(\rho, \delta), \label{Eq:meqv_cal}
\end{equation}
where $\rho_u = 2\rho_p + \rho_n = (3-\delta)\rho/2$ and $\rho_d = \rho_p + 2\rho_n= (3+\delta)\rho/2$. The interacting part of the equivalent mass can then be obtained with
\begin{equation}
  m_\mathrm{I}= m-m_{0},  \label{Eq:mI_cal}
\end{equation}
which gives the interacting energy density of nuclear matter
\begin{equation}
  E_\mathrm{I}= E(\rho_u, \rho_d, m) - E(\rho_u, \rho_d, m_0). \label{Eq:EI}
\end{equation}
Based on the obtained values for $m_\mathrm{I}$ and $E_\mathrm{I}$, the in-medium quark condensate in nuclear matter is then calculated by Eq.~(\ref{Eq:cond}).

\begin{figure}[htbp]
\centering
\includegraphics[width=\linewidth]{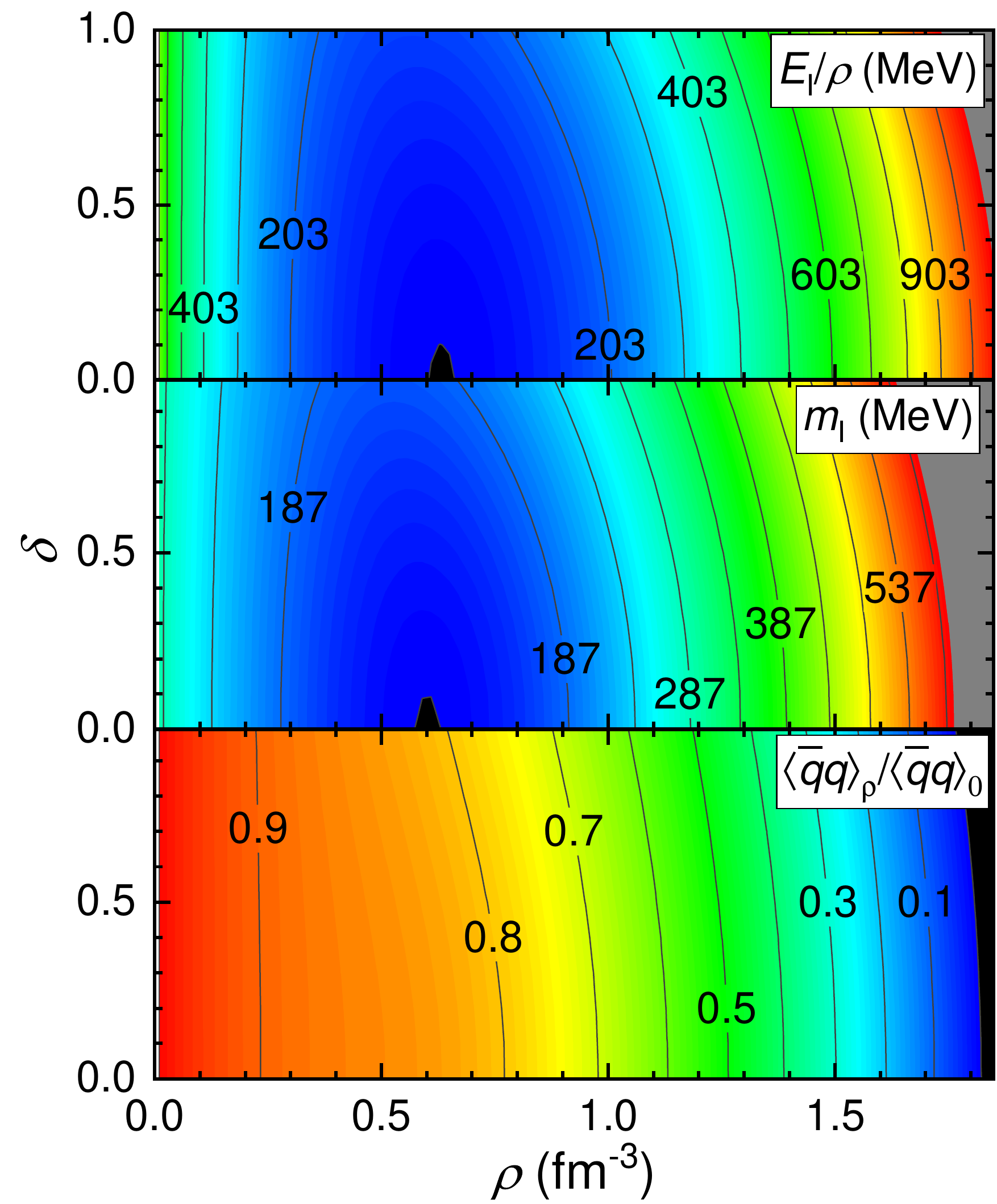}
\caption{\label{Fig:EmIcond} The interacting energy density per baryon $E_{I}/\rho$, the equivalent mass $m_\mathrm{I}$, and the relative quark condensate ${\langle\bar{q}q\rangle_{\rho}}/{\langle\bar{q}q\rangle_{0}}$ as functions of density $\rho$ and isospin asymmetry $\delta$. Typical values are adopted for the parameters in Eq.~(\ref{Eq:Et}), i.e., $\epsilon_{0}(\rho_{0})=-15.9$ MeV, $S(\rho_{0})=31.6$ MeV, $L=58.9$ MeV, $K=240$ MeV and $J=J_{sym}=K_{sym}=0$ MeV.}
\end{figure}

To give a quantitative example, adopting typical values for the parameters in Eqs.~(\ref{Eq:E0}) and (\ref{Eq:Es}), we obtain the energy density of nuclear matter in the range $0 \leq \delta \leq 1$ and $0\leq \rho \leq$ 2 fm$^{-3}$ with Eq.~(\ref{Eq:E_expand}). The equivalent mass is then determined by Eq.~(\ref{Eq:meqv_cal}), which gives the interacting parts of energy density $E_\mathrm{I}$ and equivalent mass $m_\mathrm{I}$ (as indicated in Fig.~\ref{Fig:EmIcond}) with Eqs.~(\ref{Eq:EI}) and~(\ref{Eq:mI_cal}). It is found that there exists a minimum at $\rho\approx 4\rho_0$ and $\delta=0$ for the interacting energy density per baryon $E_\mathrm{I}/\rho$, coincides with the minimum of the equivalent mass $m_\mathrm{I}$. Meanwhile, we note $m_\mathrm{I}$ increases with $\delta$ to account for the symmetry energy of nuclear matter. Based on the obtained values of $E_\mathrm{I}$ and $m_\mathrm{I}$ in Fig.~\ref{Fig:EmIcond}, the relative quark condensate ${\langle\bar{q}q\rangle_{\rho}}/{\langle\bar{q}q\rangle_{0}}$ is then fixed with Eq.~(\ref{Eq:cond}). As indicated in the lower panel of Fig.~\ref{Fig:EmIcond}, the obtained in-medium quark condensate decreases monotonically with density and finally vanishes at a rather large density $\rho \approx 1.8$ fm$^{-3}$.

\section{\label{sec:num}Results and discussions}

To constrain the properties of neutron star matter, as was done in Ref.~\cite{Zhang2018_ApJ859-90}, we carry out extensive calculations to obtain the corresponding EOSs based on the formulae introduced in Sec.~\ref{sec:the_NS}. In particular, for the parameters in Eqs.~(\ref{Eq:E0}) and (\ref{Eq:Es}), we fix $\epsilon_0(\rho_0)=-15.9$ MeV and vary the incompressibility $K$ and skewness $J$ of SNM, the slope $L$, curvature $K_{sym}$, and skewness $J_{sym}$ of the symmetry energy. Meanwhile, the symmetry energy $S(\rho_0)$ is correlated with $L$ and is fixed by Eq.~(\ref{Eq:SL}). At $\rho<0.08\ \mathrm{fm}^{-3}$, to account for the crusts of neutron stars, we adopt the EOSs presented in Refs.~\cite{Feynman1949_PR75-1561, Baym1971_ApJ170-299, Negele1973_NPA207-298}.

With the obtained EOSs for neutron star matter, the corresponding structures of neutron stars are fixed by solving the TOV equation
\begin{eqnarray}
&&\frac{\mbox{d}P}{\mbox{d}r} = -\frac{G M E}{r^2}   \frac{(1+P/E)(1+4\pi r^3 P/M)} {1-2G M/r},  \label{eq:TOV}\\
&&\frac{\mbox{d}M}{\mbox{d}r} = 4\pi E r^2, \label{eq:m_star}
\end{eqnarray}
where the gravity constant is taken as $G=6.707\times 10^{-45}\ \mathrm{MeV}^{-2}$. Meanwhile, the dimensionless tidal deformability is obtained with
\begin{equation}
\Lambda = \frac{2 k_2}{3}\left( \frac{R}{G M} \right)^5, \label{eq:td}
\end{equation}
where the second Love number $k_2$ measures how easily the star is deformed in the presence of an external tidal field and can be evaluated by introducing perturbations to the metric~\cite{Damour2009_PRD80-084035,Hinderer2010_PRD81-123016,Postnikov2010_PRD82-024016}.

\begin{table*}
\caption{\label{tablea} The saturation properties of nuclear matter corresponding to the parameters in Eqs.~(\ref{Eq:E0}) and (\ref{Eq:Es}), which are constrained from criterion (a) with the tidal deformability $\Lambda_{1.4}\leq 800$ (90\% credible region)~\cite{LVC2017_PRL119-161101}, radius $R_{1.4}=12.45\pm 1.30$ km (95\% credible region)~\cite{Miller_2021qha}, and maximum mass $M_\mathrm{TOV}< 2.05 M_{\odot}$~\cite{Cromartie2020_NA4-72}. The obtained properties of 1.4 solar-mass neutron stars (radii $R_{1.4}$, tidal deformability $\Lambda_{1.4}$, central density $\rho_{1.4}$), maximum mass $M_\mathrm{TOV}$, and maximum sound velocity $v_\mathrm{max}$ are presented as well.}
\centering
 \begin{tabular}{cccccc|cccc|c}
  \hline \hline
  $S(\rho_{0})$ &  $L$ &$K$ & $J$ & $K_\mathrm{sym}$& $J_\mathrm{sym}$  &$R_{1.4}$ & $\Lambda_{1.4}$ & $\rho_{1.4}$ & $M_\mathrm{TOV}$ &  $v_\mathrm{max}$ \\
  MeV &   MeV        &  MeV & MeV & MeV      &    MeV     &  km      &                 &   fm$^{-3}$ & $M_\odot$ &  $c$ \\  \hline
   34.9  & 80 &   220  &  200   & $-200$  & $-200$ &  11.6  &  230   &  0.77  &  1.61 & 0.81 \\
   33.8  & 70 &   240  &  200   & $-200$  & $-200$ &  11.6  &  277   &  0.70  &  1.73 & 0.91 \\
   34.9  & 80 &   240  &  200   & $-200$  & $-200$ &  12.3  &  380   &  0.60  &  1.77 & 0.88 \\
   32.7  & 60 &   260  &  200   & $-200$  & $-200$ &  11.5  &  242   &  0.66  &  1.83 & 0.97 \\
   33.8  & 70 &   260  &  200   & $-200$  & $-200$ &  12.1  &  372   &  0.59  &  1.86 & 0.95 \\
   34.9  & 80 &   260  &  200   & $-200$  & $-200$ &  12.6  &  507   &  0.51  &  1.90 & 0.93 \\
   34.9  & 80 &   220  &  0     & $-200$  &   0    &  11.6  &  251   &  0.77  &  1.61 & 0.81 \\
   33.8  & 70 &   240  &  0     & $-200$  &   0    &  11.6  &  274   &  0.70  &  1.74 & 0.91 \\
   34.9  & 80 &   240  &  0     & $-200$  &   0    &  12.2  &  419   &  0.60  &  1.78 & 0.88 \\
   32.7  & 60 &   260  &  0     & $-200$  &   0    &  11.5  &  253   &  0.66  &  1.83 & 0.97 \\
   33.8  & 70 &   260  &  0     & $-200$  &   0    &  12.1  &  395   &  0.58  &  1.87 & 0.95 \\
   34.9  & 80 &   260  &  0     & $-200$  &   0    &  12.5  &  515   &  0.51  &  1.90 & 0.93 \\
   34.9  & 80 &   240  & $-400$ & $-200$  &  600   &  12.2  &  412   &  0.60  &  1.42 & 0.50 \\
   33.8  & 70 &   260  & $-400$ & $-200$  &  600   &  12.3  &  438   &  0.52  &  1.57 & 0.55 \\
   31.6  & 50 &   240  & $-400$ & $-200$  &  800   &  11.9  &  316   &  0.52  &  1.57 & 0.59 \\
   32.7  & 60 &   240  & $-400$ & $-200$  &  800   &  12.1  &  438   &  0.51  &  1.54 & 0.56 \\
   33.8  & 70 &   240  & $-400$ & $-200$  &  800   &  12.3  &  447   &  0.51  &  1.52 & 0.54 \\
   34.9  & 80 &   240  & $-400$ & $-200$  &  800   &  12.6  &  471   &  0.50  &  1.53 & 0.53 \\
   34.9  & 80 &   220  &   0    & $-100$  & $-200$ &  12.5  &  449   &  0.52  &  1.47 & 0.51 \\
   34.9  & 80 &   220  & $-200$ & $-100$  &   0    &  12.2  &  377   &  0.62  &  1.44 & 0.48 \\
   33.8  & 70 &   240  & $-200$ & $-100$  &   0    &  12.2  &  410   &  0.56  &  1.51 & 0.50 \\
   34.9  & 80 &   260  & $-400$ & $-100$  &  200   &  12.4  &  438   &  0.59  &  1.40 & 0.46 \\
   34.9  & 80 &   240  & $-400$ & $-100$  &  400   &  12.5  &  499   &  0.56  &  1.42 & 0.47 \\
   34.9  & 80 &   260  & $-400$ & $-100$  &  400   &  12.7  &  547   &  0.48  &  1.56 & 0.50 \\
   33.8  & 70 &   240  & $-400$ & $-100$  &  600   &  12.5  &  443   &  0.49  &  1.52 & 0.51 \\
   34.9  & 80 &   240  & $-400$ & $-100$  &  600   &  12.7  &  553   &  0.48  &  1.50 & 0.50 \\
   33.8  & 70 &   260  & $-400$ & $-100$  &  600   &  12.7  &  554   &  0.45  &  1.66 & 0.53 \\
   34.9  & 80 &   260  & $-400$ & $-100$  &  600   &  12.9  &  661   &  0.44  &  1.65 & 0.52 \\
   31.6  & 50 &   240  & $-400$ & $-100$  &  800   &  12.3  &  477   &  0.46  &  1.60 & 0.55 \\
   32.7  & 60 &   240  & $-400$ & $-100$  &  800   &  12.5  &  528   &  0.45  &  1.59 & 0.54 \\
   33.8  & 70 &   240  & $-400$ & $-100$  &  800   &  12.7  &  611   &  0.45  &  1.58 & 0.52 \\
   33.8  & 70 &   240  & $-400$ &   0     &  600   &  12.9  &  603   &  0.43  &  1.57 & 0.50 \\
   34.9  & 80 &   240  & $-400$ &   0     &  600   &  13.0  &  676   &  0.43  &  1.56 & 0.49 \\
   33.8  & 70 &   240  & $-400$ &   0     &  800   &  13.0  &  630   &  0.41  &  1.60 & 0.51 \\
\hline
\end{tabular}
\end{table*}

\begin{table*}
\caption{\label{tableb} Same as Table~\ref{tablea} but with criterion (b), i.e., the tidal deformability $70\leq \Lambda_{1.4}\leq 580$ (90\% credible region)~\cite{LVC2018_PRL121-161101}, the radii $R_{1.4}=12.45\pm 0.65$ km and $R_{2.08}=12.35\pm 0.75$ km (68\% credible region)~\cite{Miller_2021qha}, and the maximum mass $M_\mathrm{TOV} \geq 2.05 M_{\odot}$ (68\% credible region)~\cite{Cromartie2020_NA4-72}. The radii of 2.08 solar-mass neutron stars $R_{2.08}$ are presented as well.}
\centering
\begin{tabular}{cccccc|ccccc|c}
  \hline  \hline
  $S(\rho_{0})$ & $L$ & $K$ & $J$ & $K_\mathrm{sym}$& $J_\mathrm{sym}$  &$R_{1.4}$ & $\Lambda_{1.4}$ &$R_{2.08}$& $\rho_\mathrm{TOV}$ & $M_\mathrm{TOV}$ & $v_\mathrm{max}$ \\
  MeV &   MeV        &  MeV & MeV & MeV      &    MeV     &  km      &           &  km      &    fm$^{-3}$ &  $M_\odot$ &  $c$ \\  \hline
 31.6 & 50 & 220 &    0   & $-300$ & 800 &  11.9 & 393 & 11.9 & 0.92 & 2.46 & 1.00 \\
 32.7 & 60 & 220 &    0   & $-300$ & 800 &  12.2 & 427 & 12.0 & 0.93 & 2.44 & 0.98 \\
 30.4 & 40 & 220 &    0   & $-200$ & 800 &  12.1 & 503 & 12.2 & 0.90 & 2.45 & 0.97 \\
 31.6 & 50 & 220 &    0   & $-200$ & 800 &  12.3 & 484 & 12.3 & 0.90 & 2.44 & 0.98 \\
 34.9 & 80 & 220 &    0   & $-100$ &  0  &  12.8 & 539 & 11.7 & 1.05 & 2.18 & 0.96 \\
 34.9 & 80 & 240 &    0   & $-100$ &  0  &  12.9 & 571 & 12.0 & 1.01 & 2.24 & 0.97 \\
 30.4 & 40 & 220 &    0   & $-100$ & 800 &  12.4 & 502 & 12.5 & 0.89 & 2.43 & 0.98 \\
 \hline
\end{tabular}
\end{table*}

The obtained mass, radius, and tidal deformability of neutron stars at various combinations of parameters are then confronted with astrophysical observations of neutron stars. More specifically, we consider two criteria:
\begin{itemize}
\item[(a)] the weaker constraints for neutron star matter at $\rho\leq\rho_{1.4}$, i.e., the tidal deformability $\Lambda_{1.4}\leq 800$ (90\% credible region)~\cite{LVC2017_PRL119-161101} and radius $R_{1.4}=12.45\pm 1.30$ km (95\% credible region)~\cite{Miller_2021qha};
\item[(b)] the stronger constraints for neutron star matter at $\rho\leq\rho_\mathrm{TOV}$, i.e., the tidal deformability $70\leq \Lambda_{1.4}\leq 580$ (90\% credible region)~\cite{LVC2018_PRL121-161101}, the radii $R_{1.4}=12.45\pm 0.65$ km and $R_{2.08}=12.35\pm 0.75$ km (68\% credible region)~\cite{Miller_2021qha}, and the maximum mass $M_\mathrm{TOV} \geq 2.05 M_{\odot}$ ($2.14^{+0.10}_{-0.09}M_{\odot}$ within 68\% credible region)~\cite{Cromartie2020_NA4-72}.
\end{itemize}
In both cases, the maximum sound velocity at $\rho\leq\rho_\mathrm{TOV}$ should not exceed the speed of light, i.e., $v_\mathrm{max}<c$. Meanwhile, to avoid double counting, we exclude the cases that fulfil the maximum mass constraint in criterion (a), i.e., $M_\mathrm{TOV}< 2.05 M_{\odot}$~\cite{Cromartie2020_NA4-72}.

\begin{figure}[!ht]
  \centering
  \includegraphics[width=\linewidth]{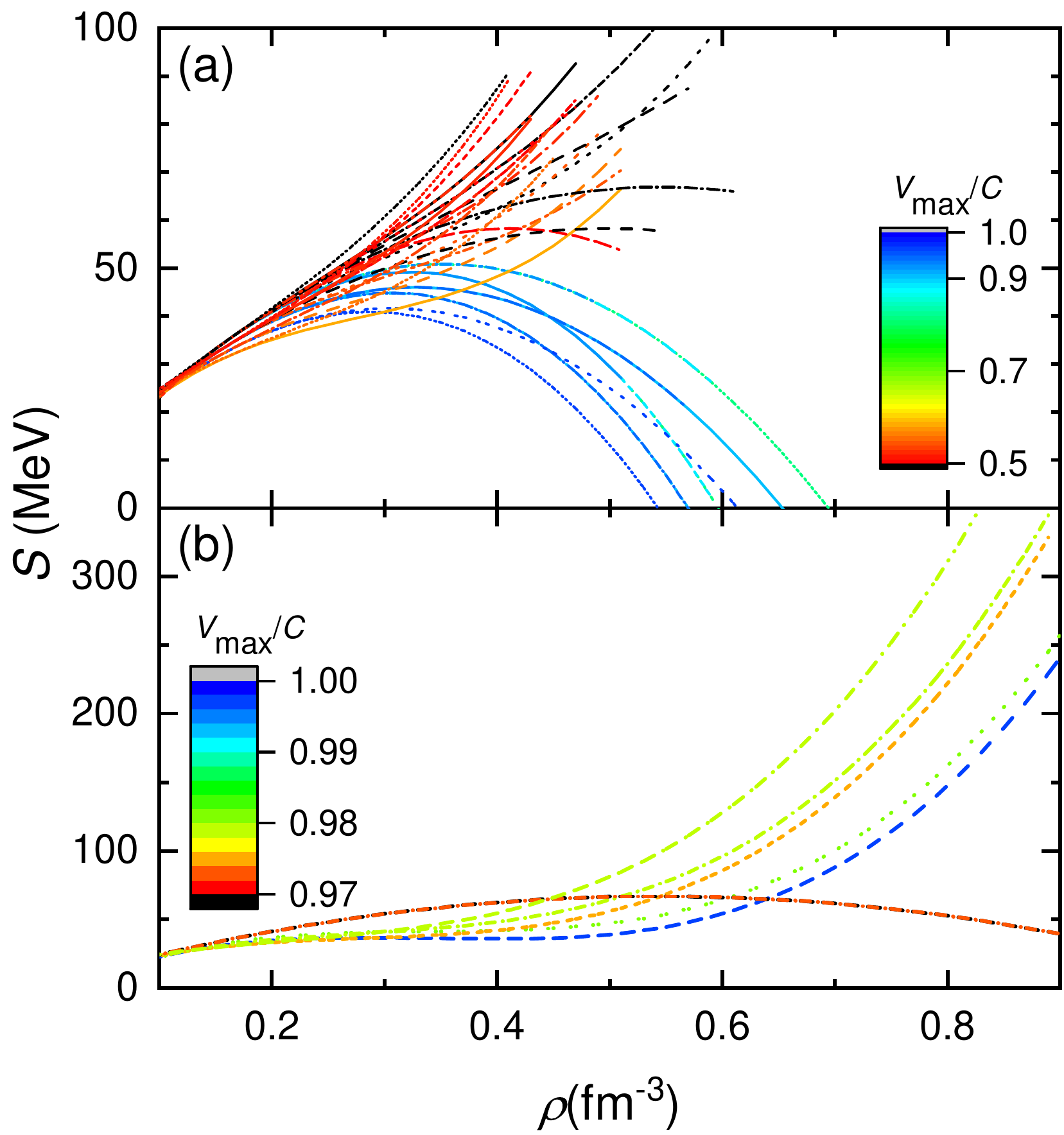}
  \caption{\label{Fig:Esym_all}The symmetry energy $S$ as functions of density $\rho$, which are determined by Eq.~(\ref{Eq:Es}) adopting the parameter sets indicated in Tables~\ref{tablea} and~\ref{tableb}. The upper panel (a) shows the results at $\rho\leq\rho_{1.4}$, where the corresponding parameters and neutron star properties are indicated in Table~\ref{tablea}. Likewise, the lower panel (b) shows the results at $\rho\leq\rho_\mathrm{TOV}$ and corresponds to Table~\ref{tableb}. The same convention is adopted for the following figures.}
\end{figure}

In Tables~\ref{tablea} and~\ref{tableb} we present the parameter sets that meet the constraints of criteria (a) and (b), as well as the corresponding neutron star properties and maximum sound velocity. Note that we have varied the parameters in the steps of 10, 20, 200, 100 and 200 MeV within the range of $L=40\sim80$ MeV, $K=220\sim260$ MeV, $J=-800\sim400$ MeV, $K_\mathrm{sym} = -400\sim100$ MeV and $J_\mathrm{sym} = -200\sim800$ MeV, respectively. For those fulfilling criterion (a), it is found that $K_\mathrm{sym}\approx -200\sim0$ MeV, which coincides with the recent constraint $K_\mathrm{sym}= -107 \pm 88$ MeV~\cite{Li2021_Universe7-182}. If criterion (b) is satisfied, we have $K_\mathrm{sym}\approx -300\sim-100$ MeV. Note that the slope of symmetry energy $L=40\sim80$ MeV can be constrained if the PREX II results with $L = 106 \pm 37$ MeV is considered~\cite{PREX2021_PRL126-172502, Reed2021_PRL126-172503}. Meanwhile, for the higher order coefficients in Eqs.~(\ref{Eq:E0}) and (\ref{Eq:Es}), criterion (a) suggests the skewness $J\approx-400\sim200$ MeV and  $J_\mathrm{sym}$ unconstrained, while criterion (b) suggests $J \approx 0$ and $J_\mathrm{sym} \approx 0\sim800$ MeV.

The constrained symmetry energy $S$ as functions of baryon number density $\rho$ are indicated in Fig.~\ref{Fig:Esym_all}, which are generally increasing with density. At the same time, we find there are few cases where $S$ is decreasing with density at $\rho\gtrsim 2\sim3\rho_0$. Particularly, the symmetry energy $S$ may even become negative at large densities adopting certain parameter sets in Table~\ref{tablea}. In such cases, the neutron star matter is comprised entirely of neutrons due to the requirement of local charge neutrality. It is interesting to note the constrained symmetry energy at two times normal density, the recent constraints from FOPI data suggests $S(2\rho_0) = 38$-73 MeV and $S(2\rho_0) = 48$-58 MeV (68\% credible region) if combined with the observational tidal deformability and maximum mass of neutron stars~\cite{Liu2021_PRC103-014616}. According to Fig.~\ref{Fig:Esym_all}, we have found $S(2\rho_0) \approx 40.1\sim65.6$ MeV for criterion (a) and $S(2\rho_0) \approx 36.4\sim 56.1$ MeV for criterion (b), which coincide with the constraints indicated in Ref.~\cite{Liu2021_PRC103-014616}.

\begin{figure}[!ht]
  \centering
  \includegraphics[width=\linewidth]{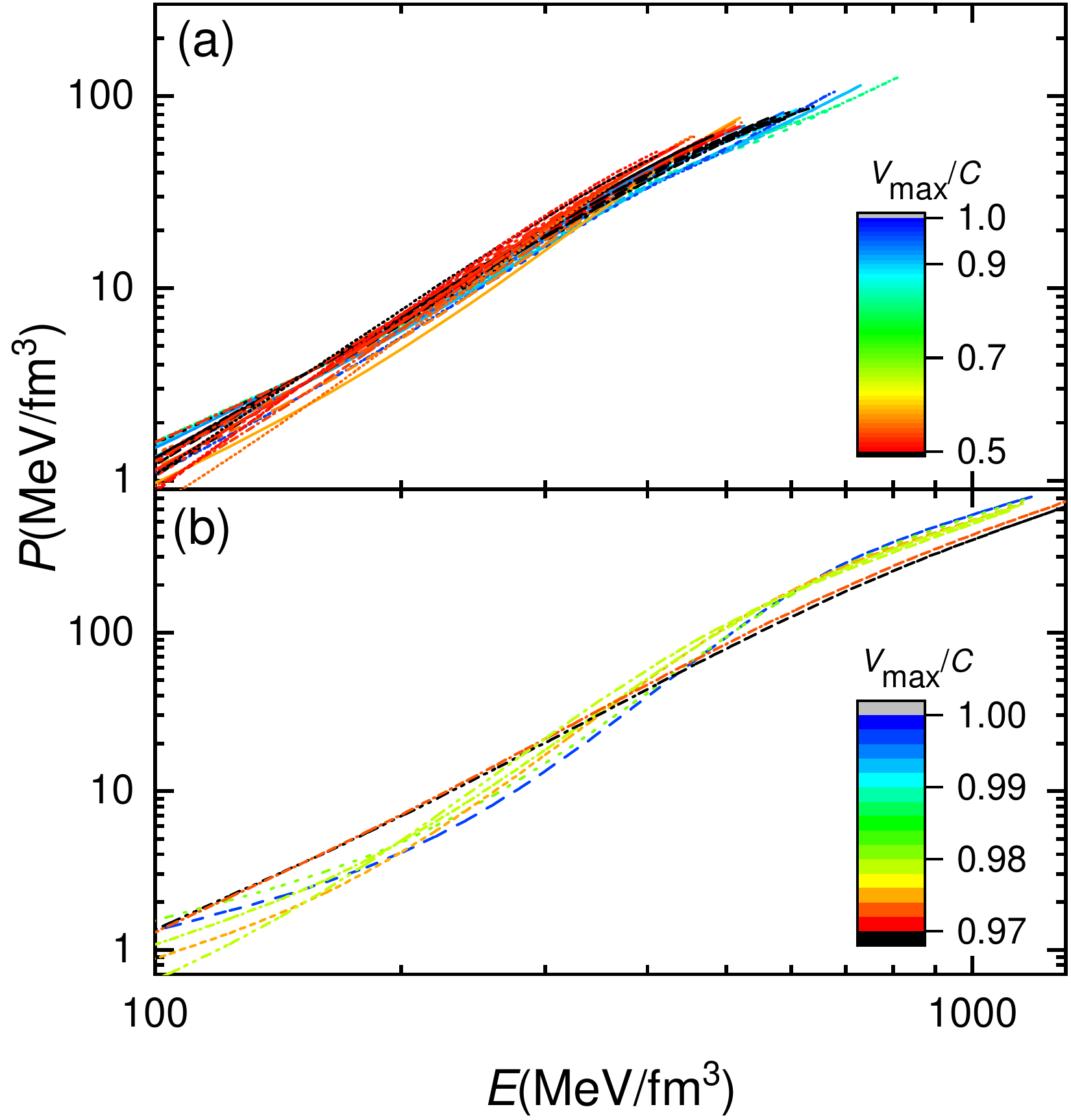}
  \caption{\label{Fig:EOS}The EOSs of neutron star matter that meet the criteria (a) and (b).}
\end{figure}

In Fig.~\ref{Fig:EOS} we present the constrained EOSs of neutron star matter, which are obtained with the parameter sets indicated in Tables~\ref{tablea} and~\ref{tableb}. Since the density of matter inside a neutron star usually dose not exceed the central density, the constrained EOSs are valid only at $\rho\leq\rho_{1.4}$ for criterion (a) and $\rho\leq\rho_\mathrm{TOV}$ for criterion (b). At larger densities, the possible emergence of new degrees of freedom can not be excluded~\cite{Sun2019_PRD99-023004}, in which case the maximum mass constraint $M_\mathrm{TOV} \geq 2.05 M_{\odot}$ (68\% credible region)~\cite{Cromartie2020_NA4-72} may be satisfied for criterion (a) if stiffer EOSs are adopted at $\rho\geq\rho_{1.4}$. In previous investigations~\cite{McLerran2019_PRL122-122701, Annala2020_NP, Xia2021_CPC45-055104}, it was shown that for the EOSs an evident deviation of the trend from lower density region is likely to take place at $E_c\approx 400$-700 MeV/fm$^{-3}$, corresponding to a maximum of sound velocity. Such a deviation is then interpreted as a phase transition from HM to QM. The EOSs indicated in the lower panel (b) of Fig.~\ref{Fig:EOS} meet the most stringent constraints from pulsar observations, where at $E\approx 200$-700 MeV/fm$^{-3}$ we have found slight deviations if large values were adopted for $K_\mathrm{sym}$ and $J_\mathrm{sym}$.

\begin{figure}[!ht]
  \centering
  \includegraphics[width=\linewidth]{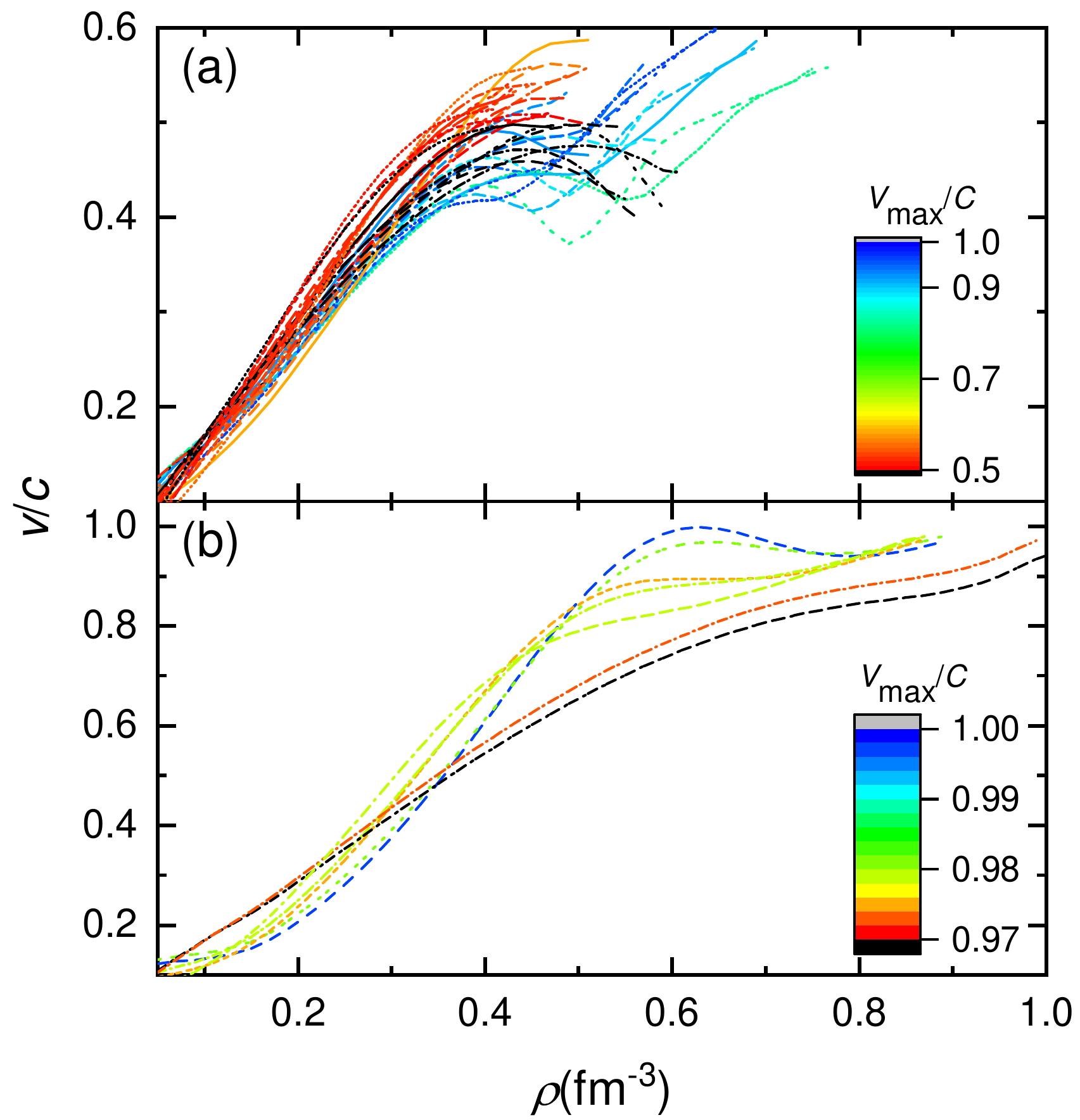}
  \caption{\label{Fig:v_all}The velocity of sound $v$ for neutron star matter obtained with the EOSs presented in Fig.~\ref{Fig:EOS}.}
\end{figure}

To show this more specifically, in Fig.~\ref{Fig:v_all} we present the velocity of sound $v$ as functions of density, which is determined by
\begin{equation}
  v = \sqrt{\frac{\mbox{d}P}{\mbox{d}E}}
\end{equation}
with the pressure $P$ and energy density $E$ indicated in Fig.~\ref{Fig:EOS}.
It is found that in most cases $v$ increases with density. However, at $\rho\approx 3\rho_0\sim5\rho_0$, there exist a maximum for the velocity of sound, coincide with the deviation of trend in the EOSs indicated in Fig.~\ref{Fig:EOS}. In particular, for a few cases fulfilling criterion (b), we find $v$ increases until reaches its peak at $v=v_\mathrm{max}\approx c$ and then decreases slightly. Meanwhile, for those with a negative symmetry energy $S(\rho)<0$, the EOSs become stiffer due to the transition into pure neutron matter, which leads to an increase of $v$ as indicated in Fig.~\ref{Fig:EOS} (a). At large densities, it is expected that $v$ approaches to the conformal limit $c/\sqrt{3}$ where a deconfinement phase transition takes place and forms QM~\cite{Annala2020_NP, Xia2021_CPC45-055104}. However, the density may be too large for neutron star matter to reach, where the maximum density $\rho_\mathrm{TOV}\lesssim 1$ fm$^{-3}$ with a rather large velocity of sound. In such cases, a full transformation from HM into a Fermi gas of quarks seems unlikely.

\begin{figure}[!ht]
  \centering
  \includegraphics[width=\linewidth]{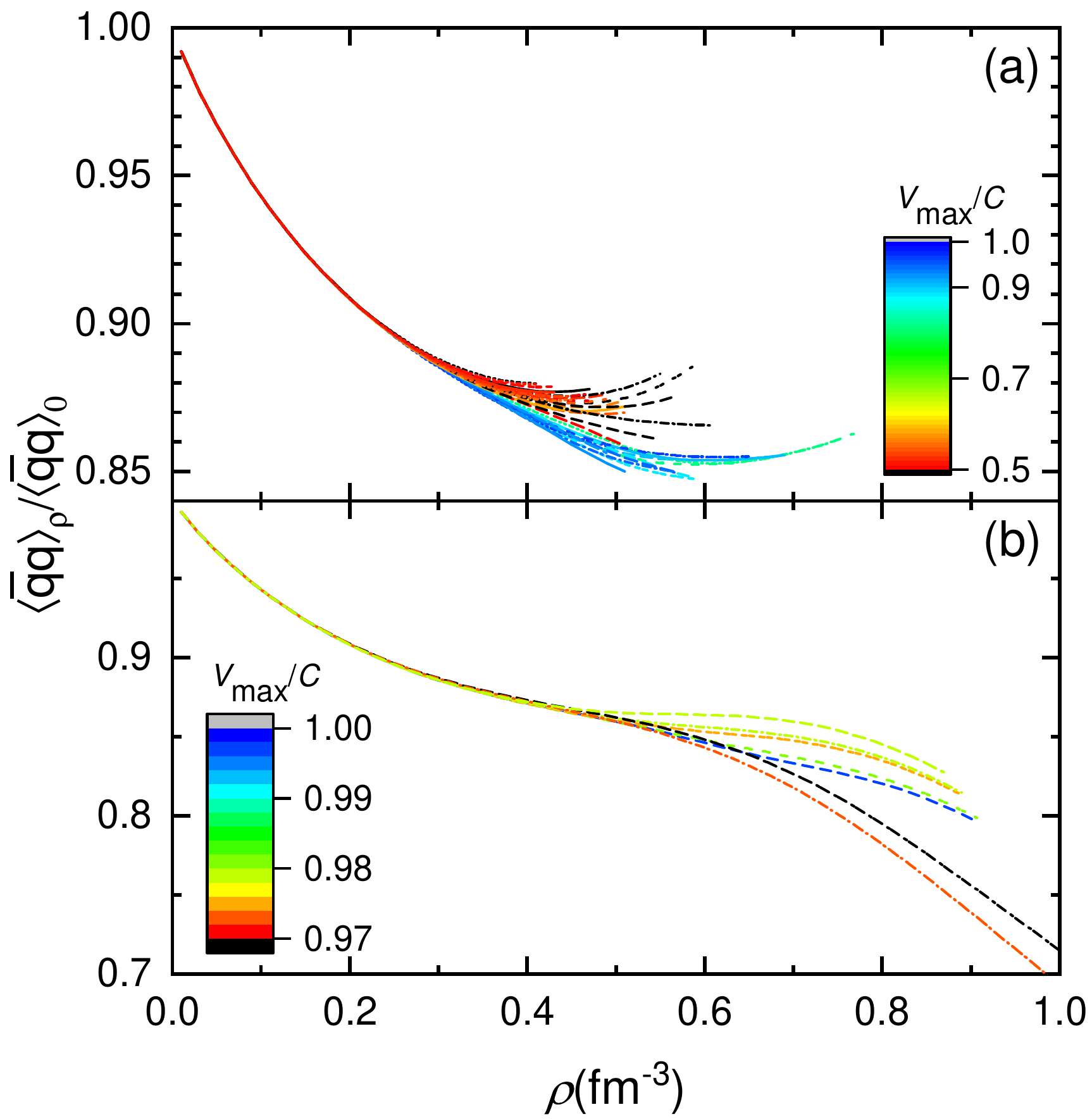}
  \caption{\label{Fig:cond_all}Relative quark condensate ${\langle\bar{q}q\rangle_{\rho}}/{\langle\bar{q}q\rangle_{0}}$ of neutron star matter as functions of baryon number density $\rho$, which correspond to the EOSs indicated in Fig.~\ref{Fig:EOS}.}
\end{figure}

Finally, based on the obtained energy density in Fig.~\ref{Fig:EOS}, the in-medium quark condensate for neutron star matter can be fixed according to the equivparticle model elaborated in Sec.~\ref{sec:the_Equiv}. The obtained results are presented in Fig.~\ref{Fig:cond_all}. We note that the relative quark condensate is generally decreasing with density, but deviate from the linear expression~\cite{Cohen1992_PRC45-1881}
\begin{equation}
 \frac{\langle\bar{q}q\rangle_\rho} {\langle\bar{q}q\rangle_0}=1-\frac{\rho}{n^*}. \label{Eq:cond_lin}
\end{equation}
At $\rho\lesssim2\rho_0$ fm$^{-3}$, the obtained quark condensates adopting various parameter sets in Tables~\ref{tablea} and~\ref{tableb} coincide with each other, which is still the case at $\rho\lesssim3\rho_0$ fm$^{-3}$ for those fulfilling criterion (b). At larger densities, the uncertainty grows due to the variations of the higher order terms in Eqs.~(\ref{Eq:E0}) and (\ref{Eq:Es}). Nevertheless, throughout the density range of neutron stars ($\rho\leq\rho_\mathrm{TOV}$), the obtained quark condensate does not vanish. In such cases, even if there are quarkyonic transition inside neutron stars, the chiral symmetry is at best partially restored.

\section{\label{sec:con}Conclusion}
In this work, we have investigated the in-medium quark condensate of neutron star matter adopting an equivparticle model~\cite{Peng2002_PLB548-189}. No extra assumptions on the current-mass derivative of model parameters are required for this approach. Exploiting the hadron-quark duality in the crossover region between HM and QM, we model nuclear matter and its transition by carrying out a Taylor expansion of the binding energy to the order of $\rho^3$~\cite{Zhang2018_ApJ859-90}. The expansion parameters are then confronted with both nuclear and astrophysical constraints. For those consistent with pulsar observations, we find the symmetry energy at large densities are consistent with various constraints from heavy-ion collisions~\cite{Liu2021_PRC103-014616}, which may even become negative at $\rho\gtrsim 3\rho_0$ if the constraints are limited to $\rho<\rho_{1.4}$. The corresponding EOSs of neutron star matter are presented as well, which may deviate slightly from the lower-density trend at $E\approx 200$-700 MeV/fm$^{-3}$ if large values of $K_\mathrm{sym}$ and $J_\mathrm{sym}$ are adopted. This deviation is further investigated with the velocity of sound $v$, which reaches its peak at $\rho\approx 3\rho_0\sim5\rho_0$. Such kind of behavior was interpreted as a phase transition from HM to QM~\cite{McLerran2019_PRL122-122701, Annala2020_NP, Xia2021_CPC45-055104}. However, we find at largest densities $\rho\approx\rho_\mathrm{TOV}$ the velocity of sound is still large and far from the conformal limit $c/\sqrt{3}$, so that a full transformation into a Fermi gas of quarks is unlikely. Based on the constrained properties of neutron star matter, we extract the corresponding quark condensate in the framework of equivparticle model~\cite{Peng2002_PLB548-189}, which is decreasing with density. It is found that the in-medium quark condensate at $\rho\lesssim2\rho_0$ fm$^{-3}$ are well constrained, while at larger densities the uncertainty grows. However, throughout the density range of neutron stars ($\rho\leq\rho_\mathrm{TOV}$), the constrained quark condensate does not vanish, which is consistent with the recent study in Ref.~\cite{Minamikawa2021}.

\section*{ACKNOWLEDGMENTS}
This work was supported by National SKA Program of China No.~2020SKA0120300, National Natural Science Foundation of China (Grant Nos.~U2032141, 11705163 \& 11875052), Ningbo Natural Science Foundation (Grant No.~2019A610066), Natural Science Foundation of Henan Province (Grant No.~202300410479), and key research projects of universities in Henan province (Grant No.~20A140003).









\end{document}